\documentclass[pre,twocolumn,reprint]{revtex4-1}

\usepackage{amsmath,amssymb}
\usepackage{bm}
\usepackage{graphicx}

\usepackage{color}

\begin{document}
\bibliographystyle{apsrev4-1}

\title{Swinging Motion of Active Deformable Particles in Poiseuille Flow}%
\author{Mitsusuke Tarama}%
\email{tarama@fukui.kyoto-u.ac.jp}%
\affiliation{
Fukui Institute for Fundamental Chemistry, Kyoto University, Kyoto, 606-8103, Japan
}%
\date{\today}%

\begin{abstract}
Dynamics of active deformable particles in an external Poiseuille flow is investigated. 
In order to make the analysis general, we employ time-evolution equations derived from symmetry considerations that take into account an elliptical shape deformation. 
First, we clarify the relation of our model to that of rigid active particles. 
Then, we study the dynamical modes that active deformable particles exhibit by changing the strength of the external flow. 
We emphasize the difference between the active particles that tend to self-propel parallel to the elliptical shape deformation and those self-propelling perpendicularly. 
In particular, a swinging motion around the centerline far from the channel walls is discussed in detail. 
\end{abstract}

\maketitle


\section{Introduction} \label{sec:Introduction}

During the last decade the dynamics of active particles has attracted much attention from the viewpoint of nonlinear and nonequilibrium science~\cite{Ramaswamy2010The,Cates2012Diffusive,Romanczuk2012Active,Marchetti2013Hydrodynamics}. 
Active particles are often found in biological systems such as microorganisms in a suspension~\cite{Lauga2009The} and the bird flocks~\cite{Vicsek2012Collective}. 
In addition, much effort has been paid to design artificial active particles. 
Among them, there are rigid active particles, which possess a prescribed time-independent shape, such as active colloids~\cite{Paxton2004Catalytic,Valadares2010Catalytic,Jiang2009Manipulation,Ebbens2010In,Bechinger2016Active}, camphor solids~\cite{Nakata1997Self-Rotation}, and bacteria~\cite{Aranson2013Active}. 
There also exist deformable ones, whose shape changes in time. 
In particular, such shape deformation is of great importance for the migration and the proliferation of eukaryotic cells~\cite{Keren2008Mechanism,Li2008Persistent,Maeda2008Ordered,Bosgraaf2009The,Mogilner2009Shape,Tanimoto2014A,Ohta2016Simple}. 
Synthetic deformable active particles are also realized such as self-propelled liquid droplets~\cite{Nagai2005Mode,Sumino2005Self-running,Toyota2009Self-propelled,Hanczyc2011Metabolism,Kitahata2011Spontaneous,Izri2014Self-Propulsion,Ebata2015Swimming,Kruger2016Curling,Yamamoto2017Chirality-induced} and active vesicles~\cite{Miura2010Autonomous,Ban2013ph-dependent,Giardini2003Compression,Boukellal2004Soft}. 

In order to elucidate the dynamics of active particles, theoretical studies of nonlinear dynamics and nonequilibrium statistical physics need to be developed. 
On the one hand, a number of elaborate models have been introduced for each specific system by taking into account the details of the internal mechanisms~\cite{Golestanian2005Propulsion,Barthes-Biesel2006From,Ishikawa2009Suspension,Kapral2013Perspective,Tjhung2012Spontaneous,Carlsson2011Mechanisms,Shao2012Coupling,Dreher2014Spiral,Ziebert2016Computational}. 
On the other hand, since the example of active particles includes both biological and synthetic systems, a basic general description of active particles is also required. 
In fact, the fundamental importance of symmetry breaking for self-propulsion is highlighted by studying simple mechanical models~\cite{Purcell1977Life,Najafi2004Simple,Gunther2008A,Alexander2008Dumb-bell}. 
For deformable active particles, we have developed a general description based on symmetry arguments and discussed the origin of basic dynamical motions~\cite{Ohta2009Deformable,Hiraiwa2010Dynamics,Hiraiwa2011Dynamics,Tarama2012Spinning,Tarama2013DynamicsPTEP,Tarama2013Oscillatory,Tarama2016Reciprocating}. 

Although active particles achieve spontaneous motion by breaking symmetry, in most practical situations this motion is influenced by the environment in various ways. 
Indeed it is an interesting problem how active particles behave as a consequence of the interplay between active motion and external stimuli such as chemical concentrations~\cite{Bagorda2008Eukaryotic,Friedrich2007Chemotaxis,Hiraiwa2013Theoretical}, light intensity~\cite{Garcia2013Light,Lozano2016Phototaxis,Jekely2009Evolution}, and mechanical forcing~\cite{Tarama2011Dynamics,Berke2008Hydrodynamic,Tailleur2009Sedimentation,Mino2011Enhanced,Elgeti2013Wall,Takagi2014Hydrodynamic,van_Teeffelen2009Clockwise-directional}. 
For active particles swimming in a fluid environment, the external flow field is crucial. 
In order to study this effect systematically, the behaviour in a characteristic flow profile, such as a linear shear flow~\cite{Tarama2013DynamicsJCP,ten_Hagen2011Brownian,Rusconi2014Bacterial}, a Poiseuille flow~\cite{Tao2010Swimming,Zottl2012Nonlinear,Zottl2013Periodic,Stark2016Swimming,Chacon2013Chaotic,Kessler1985Hydrodynamic,Mathijssen2016Upstream}, and a swirl flow~\cite{Tarama2014Deformable,Kuchler2016Getting,Sokolov2016Rapid}, is often investigated.

In our previous study~\cite{Tarama2013DynamicsJCP}, we derived model equations for active deformable particles under a general external flow field based on symmetry arguments. 
We applied the model to study the motion of active deformable particles in an external linear shear flow and compared the dynamics with those of rigid active particles. 
In addition to a straight motion and a cycloidal motion, which rigid active particles also exhibit~\cite{ten_Hagen2011Brownian}, we obtained a rich variety of dynamical modes including a winding motion, undulated cycloidal motion, and a chaotic motion due to the deformability of the particle~\cite{Tarama2013DynamicsJCP}. 

In this paper, we study the dynamics of active deformable particles in an external Poiseuille flow between two parallel walls. 
The flow profile is prescribed by a Poiseuille flow and the interaction with the confinement walls is simply modeled by a repulsive potential. 
We do not take into consideration the hydrodynamic interaction with the confinement. 
Such a case was studied before for circular and elliptical rigid active particles~\cite{Zottl2012Nonlinear,Zottl2013Periodic}, where a swinging motion around the centerline and a tumbling motion far from the centerline were identified. 
Here we are interested in how these motions are modified by the deformability of the particle shape. 

The organization of this paper is as follows.
In the next section, Sec.~\ref{section:model}, we introduce the model equations for active deformable particles in an external Poiseuille flow. 
We apply the equations of motion that have been derived for active deformable particles under a general external flow field based on symmetry considerations~\cite{Tarama2013DynamicsJCP}, to the case of a Poiseuille flow. 
First, we study the dynamics in the limit of rigid active particles in Sec.~\ref{section:rigid_poiseuille_flow}. 
The time-evolution equations are solved analytically and the dynamics of the particles of the parallel and perpendicular configurations are identified as being qualitatively the same. 
The relation of our model to that of previous works on rigid active particles in a Poiseuille flow~\cite{Zottl2012Nonlinear,Zottl2013Periodic} is also discussed. 
Then, we investigate the dynamics of active deformable particles in Sec.~\ref{section:deformable_poiseuille_flow}. 
We emphasize the difference between the deformable active particles that tend to self-propel parallel to the elliptical shape deformation and those self-propelling perpendicularly. 
In order to clarify the origin of the difference, we consider the overdamped limit in Sec.~\ref{section:overdamped_poiseuille_flow}. 
We conclude that the difference in dynamics is caused by the deformability of the particle shape. 
Section~\ref{section:discussion} is devoted to a summary and discussion.

\section{Model} \label{section:model}

In this section, we introduce the model equations of active deformable particles in a Poiseuille flow. 
For generality, we consider equations of motion derived from symmetry arguments~\cite{Tarama2013DynamicsJCP,Tarama2017Dynamics}. 
The original equations apply to both two- and three-dimensional space, but here we confine ourselves to a two-dimensional space. 

We assume that the particle is sufficiently small so that we can consider the Stokes regime. 
An external flow field is usually described by the flow velocity $\bm{u}$, and its spacial dependence is characterized by the elongation $\mathsf{A}$ and the rotation $\mathsf{W}$ defined, respectively, by
\begin{align}
&A_{ij}= \frac{1}{2} ( \partial_i u_j +\partial_j u_i ),
 \label{eq:A_ij}\\
&W_{ij} = \frac{1}{2} ( \partial_i u_j -\partial_j u_i ),
 \label{eq:W_ij}
\end{align}
where $\partial_i = d /d x_i$ denotes the spatial derivative. 
Here the subscripts $i$ and $j$ take values 1 or 2, corresponding to the $x$ and $y$ components, respectively. 
For a Poiseuille flow between two planar channel walls, which are placed at $y = \pm y_{\rm w}$, the flow velocity is prescribed as
\begin{equation}
\bm{u}(Y) = ( u_0 (1 -Y^2), 0),
 \label{eq:u}
\end{equation}
where $Y = y / y_{\rm w}$. 
$u_0$ is the maximum flow speed, which is measured at the center of the channel. 
For this flow velocity, eqs.~\eqref{eq:A_ij} and \eqref{eq:W_ij} are calculated as
\begin{align}
&\mathsf{A} = \left(
\begin{array}{cc}
0 & -\dot{\gamma}_{\rm w} Y /2 \\
-\dot{\gamma}_{\rm w} Y /2 & 0
\end{array}
\right),
 \label{eq:A}\\
&\mathsf{W} = \left(
\begin{array}{cc}
0 & \dot{\gamma}_{\rm w} Y /2 \\
-\dot{\gamma}_{\rm w} Y /2 & 0
\end{array}
\right),
 \label{eq:W}
\end{align}
where $\dot{\gamma}_{\rm w} = 2 u_0 / y_{\rm w}$ is the maximum shear rate observed at the channel wall at $Y = \pm 1$. 
\begin{figure}[tb]
\begin{center}
 \includegraphics[width=\columnwidth]{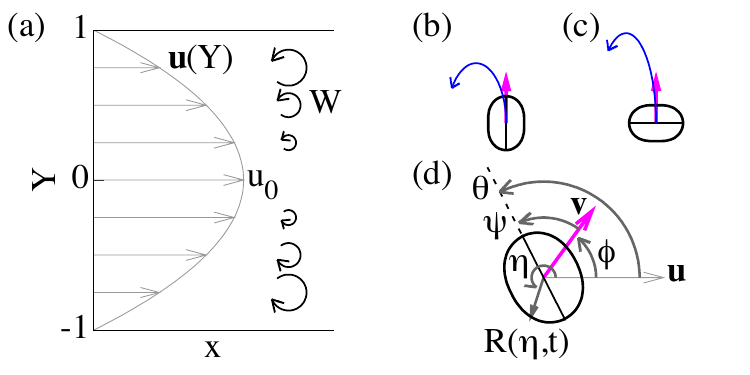}
\caption{(Colour online)
Sketches of (a) the flow velocity $\bm{u}(Y)$ and vorticity $\mathsf{W}$, the particles with (b) parallel and (c) perpendicular configurations, and 
(d) the definition of the angles $\eta$, $\phi$, $\theta$, and $\psi$.
}%
\label{fig:PoiseuilleFlow_Rigid_schematic}
\end{center}
\end{figure}
Here, note that, although the flow velocity $\bm{u}$ decreases as it goes away from the centerline and it vanishes at the channel wall $Y=\pm1$, the flow vorticity $\mathsf{W}$ increases as schematically depicted in fig.~\ref{fig:PoiseuilleFlow_Rigid_schematic}(a). 

For sufficiently small particles in Stokes flow, we decompose the centre-of-mass velocity into two contributions: 
\begin{equation}
\frac{d r_i}{dt} = u_i +v_i. 
 \label{eq:dr_i}
\end{equation}
One is the external flow velocity $\bm{u}$ representing the passive advection, and the other, denoted by $\bm{v}$, is the deviation from it. 
Here, we assume that the deviation is mainly caused by the self-propulsion. 
Therefore, we refer to $\bm{v}$ as the active velocity measured with respect to the flow velocity $\bm{u}$. 

In order to make the analysis general, we do not specify the mechanism of the active velocity. 
Instead, we consider a bifurcation structure in the equation of the active velocity. 
In addition, the effect of the particle shape as well as the vorticity of the external flow are included using symmetry considerations. 
The time-evolution equation for the active velocity is assumed to be
\begin{equation}
\frac{d v_i}{dt}
 = \alpha v_i - ( v_k v_k ) v_i -a_1 S_{ij} v_j - W_{ik} v_k + f_i. 
 \label{eq:dv_i}
\end{equation}
Here $d v_i /dt = \alpha v_i - ( v_k v_k ) v_i$ represents the supercritical pitchfork bifurcation of the active velocity, where $\alpha$ characterizes the strength of the active velocity when it is positive. 
Such a structure was also considered in a continuum description of flocks of active particles~\cite{Toner1995Long-Range,Toner1998Flocks,Toner2005Hydrodynamics}. 
The term $-a_1 S_{ij} v_j$ is the lowest-order coupling term of the active velocity and the particle shape deformation $\mathsf{S}$, and the term $- W_{ik} v_k$ represents the passive rotation due to the external flow. 
The last term $f_i$ is the interaction with the channel wall, which is specified later. 

The particle shape is described by using a traceless symmetric tensor $\mathsf{S}$, which represents an elliptical deformation (an ellipsoidal deformation in three dimensions). 
The traceless symmetric tensor is related to the local radius as follows. 
For simplicity here we consider a circular particle in a two-dimensional space. 
The local radius of the particle $R(\eta,t)$ is divided into the radius of the original circular shape $R_0$ and the deviation from it $\delta R(\eta,t)$, which is assumed to be a single-valued function of the direction $\eta$. 
The deviation can be expanded into a Fourier series as
\begin{equation}
\delta R(\eta,t) = \sum_{n=2}^{\infty} \Big( z_n(t) e^{i n \eta} + z_{-n}(t) e^{-i n \eta} \Big). 
 \label{eq:R}
\end{equation}
Here the zeroth mode, representing a uniform expansion or shrinking, is not included in eq.~\eqref{eq:R}. 
The first Fourier mode $z_{\pm1}$ induces the translation of the center of mass and, thus, is treated separately by the center-of-mass velocity $\bm{v}$. 
In general, the Fourier components can be transformed to traceless symmetric tensors. 
For example, the deformation tensor related to the second mode deformation is given by
\begin{equation}
\begin{array}{c}
S_{11} = -S_{22} = z_2 +z_{-2} = s \cos 2\theta \\
S_{12} = S_{21} = i( z_2 -z_{-2} ) = s \sin2\theta
\end{array}
 \label{eq:S}
\end{equation}
where we have defined $z_{\pm2} = (s/2) \exp(\pm 2 i \theta)$. 
$s$ and $\theta$ measure the magnitude of the elliptical deformation and the direction of the longitudinal axis, respectively. 
Then the local radius of the elliptically deformed particle is given by
\begin{equation}
R(\eta,t) = R_0 +S_{11} \cos2\eta +S_{12} \sin2\eta. 
 \label{eq:R_eta_t}
\end{equation}
The deformation tensors for higher mode deformations are defined in the same manner~\cite{Tarama2013Oscillatory}. 
However, throughout this paper, we only consider the second mode deformation $\mathsf{S}$, which represents an elliptical shape in two dimensions. 
Similarly, the deformation tensors in three dimensions are related to the coefficients of the spherical harmonic expansion of the local radius~\cite{Hiraiwa2011Dynamics,Fel1995Tetrahedral}. 

Based on the symmetry considerations, the time-evolution equation for the elliptical deformation tensor $\mathsf{S}$ is derived as 
\begin{align}
&\frac{d S_{ij}}{dt}
 = - \Big( \kappa + \frac{1}{2}S_{k\ell} S_{k\ell} \Big) S_{ij} + b_1 \Big( v_i v_j -\frac{v_k v_k}{2} \delta_{ij} \Big) \notag\\
 & \hspace{1.2em}- ( W_{ik} S_{kj} +W_{jk} S_{ki} ) 
 + \nu_1 \Big( A_{ij} -\frac{1}{2} A_{kk} \delta_{ij} \Big).
 \label{eq:dS_ij}
\end{align}
where $\delta_{ij}$ denotes the Kronecker delta. 
Here we assume that the shape deformation does not occur spontaneously and therefore we keep $\kappa>0$, 
which corresponds to the stiffness of the deformation. 
The term with the coefficient $b_1$ is the lowest coupling term between the active velocity and the elliptical shape deformation. 
This term is the counterpart of the term with $a_1$ in eq.~\eqref{eq:dv_i} but the sign is set to be opposite to make the system nonvariational~\cite{Tarama2014Individual}. 
The term $- ( W_{ik} S_{kj} +W_{jk} S_{ki} )$ represents the passive rotation due to the external flow, which is simultaneously stretching the particle through the $\nu_1$ term.

\begin{figure}[tb]
\begin{center}
 \includegraphics[width=\columnwidth]{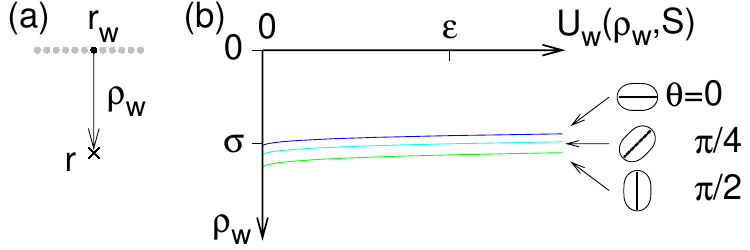}
\caption{(Colour online)
The interaction potential with the confinement. 
(a) The actual relative position $\bm{\rho}_{\rm w}$ of the particle at $\bm{r}$ measured from the element of the confinement at $\bm{r}_{\rm w}$. 
The confinement is composed of such small circular elements. 
(b) The potential $U_{\rm w}$ as a function of the actual relative distance $\rho_{\rm w}$ for different configurations $\mathsf{S}$ with $\theta=0$, $\pi/4$, and $\pi/2$. 
The relative particle position and the magnitude of the deformation is set to $\bm{\rho}_{\rm w}=(0,-\rho_{\rm w})$ and $s=0.2$, respectively. 
}%
\label{fig:PoiseuilleFlow_potential_schematic}
\end{center}
\end{figure}

Finally, the effect of the confinement is included by a shifted and truncated Lennard-Jones potential
\begin{equation}
\bar{U}_{\rm{w}} = 
\int d r_{\rm w} U_{\rm{w}},
 \label{eq:U_w}
\end{equation}
where the integral runs over the confinement and
\begin{equation}
U_{\rm{w}} = 
 4 \epsilon \Big[ 
 \Big( \frac{\sigma}{ \tilde{\rho}_{\rm w} } \Big)^{12} 
 - \Big( \frac{\sigma}{ \tilde{\rho}_{\rm w} } \Big)^{6} 
 +\frac{1}{4}\Big]  H( 2^{1/6} \sigma -\tilde{\rho}_{\rm w} ).
 \label{eq:U_w_element}
\end{equation}
$H(x)$ is the Heaviside step function, which gives 1 if $x>0$ and 0 otherwise. 
Here we assume that the confinement is composed of small circular elements, 
with each of which the elliptically deformable particle interacts through the potential given by eq.~\eqref{eq:U_w_element}. 
The effect of the particle deformation is included into the distance between the particle and the element of the confinement as
\begin{equation}
\tilde{\rho}_{\rm w} = \big( \bm{\rho}_{\rm w} \cdot ( \mathsf{I} -\mathsf{S} )\cdot \bm{\rho}_{\rm w} \big)^{1/2}, 
 \label{eq:tilde_rho}
\end{equation}
where $\mathsf{I}$ is the identity matrix and 
$\bm{\rho}_{\rm w} = \bm{r} -\bm{r}_{\rm w}$
 is the actual relative position of the particle measured from the confinement element at $\bm{r}_{\rm w}$. 
See also the sketches in fig.~\ref{fig:PoiseuilleFlow_potential_schematic}. 
Such effective distance was also considered in Ref.~\cite{Menzel2012Soft}, where deformable active particles with soft repulsive interactions were studied. 
Note that the potential in eq.~\eqref{eq:U_w_element} is introduced in the same spirit as the Gay-Berne potential~\cite{Gay1981Modification} in the limit that one particle is circular. 
Then, the confinement force on the particle velocity is calculated from eq.~\eqref{eq:U_w} as
\begin{align}
&f_i = -\frac{\partial \bar{U}_{\rm w}}{\partial \bm{r}} 
 = \int d r_{\rm w} \frac{24 \epsilon}{\sigma^2} (\delta_{ij} -S_{ij} ) \tilde{\rho}_{{\rm w} j} H( 2^{1/6} \sigma -\tilde{\rho}_{\rm w} )  \notag\\
& \hspace{9em} \times \Big[  2  \Big( \frac{\sigma}{ \tilde{\rho}_{\rm w} } \Big)^{14} 
 - \Big( \frac{\sigma}{ \tilde{\rho}_{\rm w} } \Big)^{8}  \Big].
 \label{eq:f_i}
\end{align}
For simplicity we here do not consider the contribution to the shape deformation in eq.~\eqref{eq:dS_ij}.

In this paper we do not consider active rotation of the particle~\cite{Tarama2013DynamicsJCP,Tarama2012Spinning,Tarama2013DynamicsPTEP,Wittkowski2012Self-propelled}. 
We also note that the hydrodynamic interaction with the confinement is not included in eqs.~\eqref{eq:dr_i}, \eqref{eq:dv_i}, and \eqref{eq:dS_ij}, which plays an important role for, i.e., the particle-particle interaction and the interaction between particles and confinement. 
Therefore, the omission of the hydrodynamic interaction is justified for the single particle dynamics far from the confining walls. 
Throughout this paper, we only discuss the single particle dynamics.

In the absence of the external flow and confinement, it is known that the model equations \eqref{eq:dv_i} and \eqref{eq:dS_ij} exhibits a bifurcation from a straight motion, where the particles self-propel in a straight line, to a circular motion, where they draw a circular trajectory, for $\alpha>0$ and $\kappa>0$~\cite{Ohta2009Deformable,Hiraiwa2010Dynamics}. 
We note that, in this limit, the equations~\eqref{eq:dv_i} and \eqref{eq:dS_ij} have been derived around the drift bifurcation in two different systems. 
On the one hand, it was derived for the isolated domain solution in reaction-diffusion equations~\cite{Ohta2009Deformation,Shitara2011Deformable}. 
On the other hand, it is also derived from the Stokes equation for a self-propelled droplet catalyzing a chemical reaction on its interface that changes the interfacial tension~\cite{Yabunaka2012Self-propelled,Yoshinaga2014Spontaneous}. 
The time-derivative term of the center-of-mass velocity arises as a consequence of the time delay, that is, the relaxation of the chemical concentration is much slower than the fluid velocity.

To conclude this section, we rewrite eqs.~\eqref{eq:dr_i}, \eqref{eq:dv_i}, and \eqref{eq:dS_ij} in a different form. 
We measure the particle center-of-mass position and velocity by 
\begin{align}
&\bm{x} = ( x, y ),
 \label{eq:x}\\
&\bm{v} = ( v \cos\phi, v\sin\phi ),
 \label{eq:v}
\end{align}
and the shape deformation by eq.~\eqref{eq:S}. 
Here, note that the angles $\phi$ in eq.~\eqref{eq:v} and $\theta$ in eq.~\eqref{eq:S} measure the orientation with respect to the flow direction.
Therefore, $\phi=0$ corresponds to the downstream orientation and $\phi=\pi$ to the upstream orientation, respectively. 
In the same manner, $\theta=0$ represents the elliptical deformation parallel to the streamline, while $\theta=\pi/2$ denotes the elongation perpendicular to the flow velocity. 
See the sketch in fig.~\ref{fig:PoiseuilleFlow_Rigid_schematic}(d) for the definition of the angles. 
Then eqs.~\eqref{eq:dr_i}, \eqref{eq:dv_i}, and \eqref{eq:dS_ij} become
\begin{align}
&\frac{d X}{d t} =  \frac{v}{y_{\rm w}} \cos\phi +  \frac{\dot{\gamma}}{2} \Big( 1 - Y^2 \Big),
 \label{eq:dx}\\
&\frac{d Y}{d t} = \frac{v}{y_{\rm w}} \sin\phi,
 \label{eq:dy}\\
&\frac{d v}{d t} = \alpha v -v^3 -a_1 v s \cos2\psi +\big( f_1 \cos\phi +f_2 \sin\phi \big),
 \label{eq:dv}\\
&\frac{d \phi}{d t} = -a_1 s \sin2\psi + \frac{\dot{\gamma}_{\rm w}}{2} Y -\frac{1}{v} \big( f_1 \sin\phi -f_2 \cos\phi \big),
 \label{eq:dphi}\\
&\frac{d s}{d t} = - \kappa s +\frac{b_1}{2} v^2 \cos2\psi -\nu_1 \frac{\dot{\gamma}_{\rm w}}{2} Y \sin2\theta,
 \label{eq:ds}\\
&\frac{d \theta}{d t} = \frac{1}{2s} \Big[ -\frac{b_1}{2} v^2 \sin2\psi + \dot{\gamma}_{\rm w} Y s -\nu_1 \frac{\dot{\gamma}_{\rm w}}{2} Y \cos2\theta \Big],
 \label{eq:dtheta}
\end{align}
where we write $\psi = \theta -\phi$. 
From eqs.~\eqref{eq:dphi} and \eqref{eq:dtheta}, we obtain the equation for $\psi$ as
\begin{align}
\frac{d \psi}{dt}
 &= \frac{1}{2 s} \Big[ \Big( 2 a_1 s^2 -\frac{b_1}{2} v^2 \Big) \sin 2\psi -\nu_1 \frac{\dot{\gamma}_{\rm w}}{2} Y \cos2\theta \Big] \notag\\
 &+ \frac{1}{v} \big( f_1 \sin\phi -f_2 \cos\phi \big).
 \label{eq:dpsi}
\end{align}
Note that $X$ in eq.~\eqref{eq:dx} is defined in the same way as $Y = y / y_{\rm w}$, i.e., $X = x /y_{\rm w}$.

\section{Rigid active particles} \label{section:rigid_poiseuille_flow}

In this section, we consider the case of rigid active particles in the Poiseuille flow. 
For this purpose, we replace eqs.~\eqref{eq:dv} and \eqref{eq:ds} by constant values for the active velocity and shape deformation, as explained shortly. 
In order to highlight the effect of the Poiseuille flow, we do not consider the interaction with the boundary explicitly in this section.
That is, we set $f_1 = f_2 = 0$ and restrict the particle position to $-1<Y<1$.

In the case of a rigid elliptical particle, we can replace eq.~\eqref{eq:ds} by
\begin{equation}
s = s_0,
 \label{eq:s0}
\end{equation}
where $s_0$ is a non-negative constant describing the magnitude of the prescribed elongation.
For simplicity, we further assume that the active velocity is a constant, which corresponds to the overdamped limit, and replace eq.~\eqref{eq:dv} by 
\begin{equation}
v = v_0.
 \label{eq:v0}
\end{equation}
Here the dependence of the active velocity on the deformation $s$ and $\theta$ represented by the $a_1$ term in eq.~\eqref{eq:dv} is also omitted. 

Now, we eliminate the time-evolution equation for $\psi$ by setting $d\psi/dt=0$ in eq.~\eqref{eq:dpsi}, yielding 
\begin{equation}
\sin2\psi = \frac{\nu_1 \dot{\gamma}_{\rm w} Y}{4a_1 s_0^2 -b_1 v_0^2} \cos2\theta.
 \label{eq:sin2psi}
\end{equation}
Then, both eqs.~\eqref{eq:dphi} and \eqref{eq:dtheta} become
\begin{equation}
\frac{d \phi}{dt} =\frac{d \theta}{dt} = \frac{\dot{\gamma}_{\rm w}}{2} Y ( 1 -G \cos2\theta ),
 \label{eq:dphi=dtheta}
\end{equation}
where the geometric factor
\begin{equation}
G =   \frac{2 \nu_1 a_1 s_0}{4 a_1 s_0^2 -b_1 v_0^2} 
 \label{eq:rigid:g}
\end{equation}
characterizes the particle shape. 
The circular shape is represented by $G=0$. 

Considering the symmetry, there are two situations concerning the relation between the directions of the active velocity and the particle shape. 
One is the parallel configuration, where the active velocity is fixed to the longitudinal axis, i.e., $\phi=\theta$, as depicted in fig.~\ref{fig:PoiseuilleFlow_Rigid_schematic}(b). 
The other is the perpendicular configuration, where the particle self-propel in the direction perpendicular to the longitudinal axis $\phi=\theta\mp\pi/2$, as illustrated in fig.~\ref{fig:PoiseuilleFlow_Rigid_schematic}(c). 
Note that, from eq.~\eqref{eq:dphi=dtheta}, particles moving upwards with the parallel configuration are turned stronger than those with the perpendicular configuration, as sketched in figs.~\ref{fig:PoiseuilleFlow_Rigid_schematic}(b) and (c) for $Y>0$. 

Now we integrate eqs.~\eqref{eq:dy} and \eqref{eq:dphi=dtheta} analytically. 
Afterwards we can solve eq.~\eqref{eq:dx}. 
However, here we do not consider the explicit formula of the solution of eq.~\eqref{eq:dx}. 
Note that there is a translational symmetry in $x$ direction and thus, eqs.~\eqref{eq:dy} and \eqref{eq:dphi=dtheta} are independent of $X$. 

In the case of the parallel configuration, we can identify the constant of motion of eqs.~\eqref{eq:dy} and \eqref{eq:dphi=dtheta} as
\begin{equation}
C_0 = \frac{1}{4} \frac{\dot{\gamma}_{\rm w} y_{\rm w}}{v_0} Y^2 +\frac{\tanh^{-1} [ \sqrt{2 G /(1+G)} \cos\phi ]}{ \sqrt{2 G (1+G)} }.
 \label{eq:rigid:C0_para}
\end{equation}
On the other hand, for the perpendicular configuration, it becomes
\begin{equation}
C_0 = \frac{1}{4} \frac{\dot{\gamma}_{\rm w} y_{\rm w}}{v_0} Y^2 +\frac{\tan^{-1} [ \sqrt{2 G /(1-G)} \cos\phi ]}{ \sqrt{2 G (1-G)} }.
 \label{eq:rigid:C0_perp}
\end{equation}
Here, $\tanh^{-1}$ and $\tan^{-1}$ denote the inverse functions of the hyperbolic tangent and that of the tangent, respectively. 
Besides, in the limit of a circular particle $G=0$, both eqs.~\eqref{eq:rigid:C0_para} and \eqref{eq:rigid:C0_perp} are reduced to 
\begin{equation}
C_0 = \frac{1}{4} \frac{\dot{\gamma}_{\rm w} y_{\rm w}}{v_0} Y^2 +\cos \phi.
 \label{eq:rigid:C0}
\end{equation}
In each case, the solutions of eqs.~\eqref{eq:dy} and \eqref{eq:dphi=dtheta} are determined by the above constant of motion. 
Different values of $C_0$ describe different solutions, which are marginally stable. 

The constants of motion, eqs.~\eqref{eq:rigid:C0_para} and \eqref{eq:rigid:C0}, are consistent with those obtained by Z\"ottl and Stark~\cite{Zottl2012Nonlinear,Zottl2013Periodic}, who studied the motion of an elliptical rigid particle swimming along the longitudinal axis, as well as circular-shaped rigid active particles, in a Poiseuille flow. 
In fact, the set of equations \eqref{eq:dy} and \eqref{eq:dphi=dtheta} in the case of the parallel configuration ($\phi=\theta$) have the same form as those studied in Ref.~\cite{Zottl2012Nonlinear,Zottl2013Periodic}. 
They found two solutions, which correspond to swinging motion around the centerline and tumbling motion. 
They also showed numerically that the swinging swimmer achieved a net-upstream motion for a sufficiently small flow velocity, while it swam downstream for a large flow velocity. 

\begin{figure}[tb]
\begin{center}
 \includegraphics[width=\columnwidth]{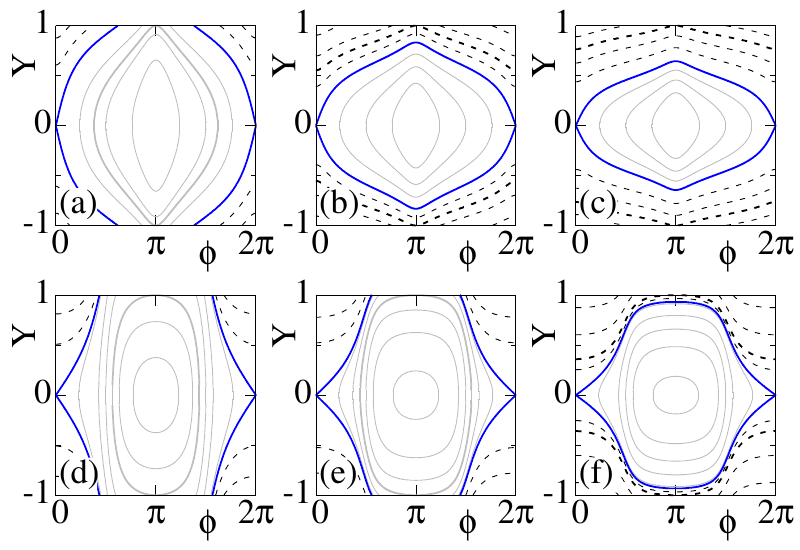}
\caption{(Colour online)
The trajectories in the $Y$-$\phi$ space of the solutions of rigid active particles with [(a)--(c)] the parallel configuration and [(d)--(f)] the perpendicular configuration. 
The gray solid lines and the black broken lines correspond to the swinging motion around the centerline and the tumbling motions, respectively, which are separated by the separatrix displayed by the blue (dark grey) solid line. 
In each panel, different curves correspond to different values of $C_0$. 
The flow velocity is chosen as $u_0 = \dot{\gamma}_{\rm w} y_{\rm w}/2 = 0.25$ for (a) and (d), 0.6 for (b) and (e), and 1 for (c) and (f). 
The other parameters are fixed as $v_0 = 0.1$ and $G=0.8$. 
The results of the parallel configuration displayed in panels (a)--(c) are consistent with those in Ref.~\cite{Zottl2013Periodic}. 
}%
\label{fig:PoiseuilleFlow_Rigid}
\end{center}
\end{figure}

Here, we consider the difference between the elliptical rigid active particles of the parallel and perpendicular configurations. 
The solutions obtained from Eqs.~\eqref{eq:rigid:C0_para} and \eqref{eq:rigid:C0_perp} 
are plotted in $Y$-$\phi$ space for different strength of the flow velocity $u_0$ in fig.~\ref{fig:PoiseuilleFlow_Rigid}, where the trajectories for the parallel and perpendicular configurations are displayed in the panels (a)--(c) and (d)--(f), respectively. 
The flow velocity is set as $u_0=0.25$ for (a) and (d), $=0.6$ for (b) and (e), and $=1$ for (c) and (f), whereas the other parameters are fixed as $v_0 = 0.1$ and $G=0.8$.
The gray solid lines correspond to the swinging motion around the centerline, whereas the black broken lines represent the tumbling motions. 
These solutions are divided by a separatrix displayed by the blue (dark grey) solid line. 
The size of the swinging trajectory increases with $C_0$. 
Upon a further increase in $C_0$, it becomes larger than the separatrix and the tumbling motion appears. 
These results indicate that the dynamics of the parallel and perpendicular configurations are qualitatively the same. 

In both parallel and perpendicular configurations, the time-evolution direction of the trajectories in $Y$-$\phi$ space for the swinging motion depicted in fig.~\ref{fig:PoiseuilleFlow_Rigid} is always clockwise. 
This is easily understood as follows. 
When the particle is swimming upwards ($0<\phi<\pi$) in the upper half plane $0<Y<1$, it suffers from the external vorticity $\mathsf{W}$, which rotates the active velocity $\bm{v}$ counterclockwise and, therefore, the angle $\phi$ increases until it crosses $\pi$. 
Then the particle swims downwards ($\pi<\phi<2\pi$) and enters to the lower half plane $-1<Y<0$, where the vorticity $\mathsf{W}$ rotates $\bm{v}$ clockwise, and thus $\phi$ decreases until it crosses $\pi$ and the particle swims upwards. 
As a result, the swinging motion corresponds to a clockwise rotation around $(\phi,Y)=(\pi,0)$ in $Y$-$\phi$ space. 
Here note that, while the flow profile \eqref{eq:u} is symmetric with respect to $Y=0$, the vorticity $\mathsf{W}$ is antisymmetric.

\section{Deformable active particles} \label{section:deformable_poiseuille_flow}

Now we consider the dynamics of deformable active particles. 
Since it is difficult to solve the full equations, eqs.~\eqref{eq:dr_i}, \eqref{eq:dv_i}, and \eqref{eq:dS_ij}, or equivalently eqs.~\eqref{eq:dx}--\eqref{eq:dtheta}, analytically, we integrate the time-evolution equations numerically
 including the interaction with the walls $\bm{f}$. 
The fourth-order Runge-Kutta method with the time increment $\delta t=10^{-3}$ is employed. 
Again we distinguish the particles that tend to self-propel in the parallel direction with respect to the longitudinal axis of the elliptical shape deformation ($a_1=b_1=1$) and those self-propelling perpendicularly ($a_1=b_1=-1$). 
Hereafter, we refer to the former as the parallel particles and the latter as the perpendicular particles. 
In both cases, we vary the self-propulsion strength $\alpha$ as well as the external flow speed $u_0$, which is measured by the maximum local shear rate $\dot{\gamma}_{\rm w} = 2 u_0/ y_{\rm w}$. 
The other parameters are fixed as $\kappa=0.5$, $\nu_1=1$, 
$y_{\rm w}=20$, $\sigma=\epsilon=1$. 

The results for the parallel and perpendicular particles are summarized in figs.~\ref{fig:PoiseuilleFlow_Wall_parallel} and \ref{fig:PoiseuilleFlow_Wall_perpendicular}, respectively. 
In both cases, the dynamical phase diagram is displayed in panel (a), whereas in panels (b)--(f) characteristic trajectories in real space as well as those in $Y$-$\phi$ and $Y$-$\theta$ spaces are plotted. 
In the real space plots the periodic boundary condition is applied in the $x$ direction for a better illustration, where the window width is set to 
$X=2.5$ (i.e., $x=50$). 
Some particle silhouettes are superposed onto the center-of-mass trajectories together with the active velocity (magenta vectors) and the longitudinal axis of the elliptical deformation (black bars). 
Note that the flow profile of eq.~\eqref{eq:u} is symmetric with respect to $Y=0$, while its spacial derivatives $\mathsf{A}$ and $\mathsf{W}$ in eqs.~\eqref{eq:A} and \eqref{eq:W} are antisymmetric. 
Therefore, the trajectories in real space is symmetric with respect to $Y=0$, whereas those in $Y$-$\phi$ space and $Y$-$\theta$ space are symmetric with respect to $(Y,\phi)=(0,\pi)$ and $(Y,\theta)=(0,0)$, respectively. 

In the following, we explain the dynamics of the parallel and perpendicular particles one by one in detail. 
We note that, in the absence of the external flow field and the confinement, the particles exhibit a straight motion for $0<\alpha<\alpha_c$ and a circular motion for $\alpha>\alpha_c$~\cite{Ohta2009Deformable}.

\subsection{Parallel particles} \label{section:parallel_poiseuille_flow}

First we show the results for the parallel particles, which tend to self-propel in the parallel direction with respect to the longitudinal axis of the elliptical shape deformation ($a_1=b_1=1$). 
As we will see in the following, the dynamics observed for $0<\alpha<\alpha_c$ are similar to those of the rigid active particles in the Poiseuille flow discussed in Sec.~\ref{section:rigid_poiseuille_flow}. 

\begin{figure*}[tb]
\begin{center}
 \includegraphics[width=\textwidth]{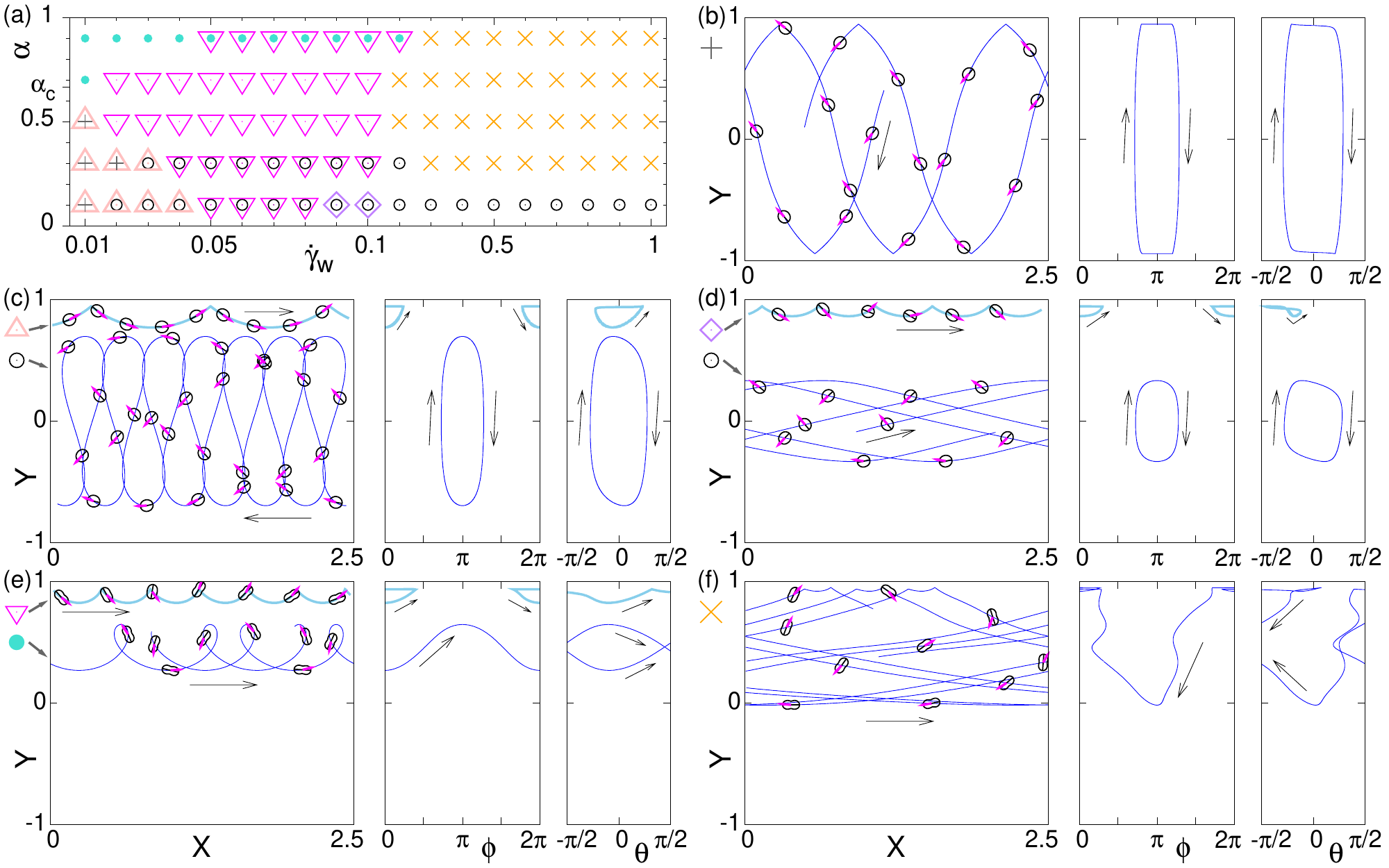}
\caption{(Colour online)
Dynamics of parallel particles ($a_1=b_1=1$) in the Poiseuille flow. 
Different dynamical states are denoted by different symbols in panel (a), which are plotted in panels (b)--(f) consistently. 
(a) Dynamical phase diagram for varying $\alpha$ and $\dot{\gamma}_{\rm w}$. 
The superposition of the symbols indicate coexistence. 
Note that the scale of the $x$ axis is different for $\dot{\gamma}_{\rm w}<0.1$ and for $\dot{\gamma}_{\rm w}>0.1$. 
(b)--(f) Characteristic trajectories of each dynamical motion in real space (left), in $Y$-$\phi$ space (middle), and in $Y$-$\theta$ space (right). 
(b) Bouncing motion between the two channel walls for $\alpha=0.1$ and $\dot{\gamma}_{\rm w}=0.01$;
(c) Upstream swinging motion [blue (dark grey) lines] and bouncing motion against the wall with the shape oscillating around $\theta=0$ [sky blue (light grey) lines] for $\alpha=0.1$ and $\dot{\gamma}_{\rm w}=0.02$;
(d) Downstream swinging motion [blue (dark grey) lines] and bouncing motion against the wall with the shape oscillating around the diagonal direction [sky blue (light grey) lines] for $\alpha=0.1$ and $\dot{\gamma}_{\rm w}=0.09$;
(e) Cycloidal motion [blue (dark grey) lines] and tumbling motion at the wall [sky blue (light grey) lines] for $\alpha=0.9$ and $\dot{\gamma}_{\rm w}=0.06$;
(f) Bouncing with excursion to the middle of the channel for $\alpha=0.9$ and $\dot{\gamma}_{\rm w}=0.4$. 
In panels (b)--(f), some particle silhouettes are superposed onto the real space trajectories with the magenta arrows and the black bars indicating the direction of the self-propulsion $\bm{v}$ and the longitudinal axis of the elliptical deformation, respectively. 
Black arrows indicate the time-evolution direction. 
}%
\label{fig:PoiseuilleFlow_Wall_parallel}
\end{center}
\end{figure*}

For $0<\alpha<\alpha_c$, the particles undergo a straight motion when the external flow field is absent. 
When the external Poiseuille flow exists, a swinging motion around the centerline is obtained. 
The parameters where the swinging motion is observed are denoted by the black circles in fig.~\ref{fig:PoiseuilleFlow_Wall_parallel}(a). 
A characteristic trajectories of the swinging motion is depicted by the blue lines in fig.~\ref{fig:PoiseuilleFlow_Wall_parallel}(d) for $\alpha=0.1$ and $\dot{\gamma}_{\rm w}=0.04$. 
The active velocity $\bm{v}$ of the swinging motion of the active deformable particles tends to point upstream and its direction $\phi$ oscillates around $\pi$. 
Therefore, since the particle motion in $x$ direction is determined by the superposition of the self-propulsion $\bm{v}$ and the passive advection due to the external flow $\bm{u}$ as in eq.~\eqref{eq:dx}, the particle may drift upstream if the flow velocity is sufficiently small. 
Indeed, an upstream swinging motion is observed for small $\dot{\gamma}_{\rm w}$ as displayed by the blue lines in fig.~\ref{fig:PoiseuilleFlow_Wall_parallel}(c) for $\alpha=0.1$ and $\dot{\gamma}_{\rm w}=0.02$. 
Note that the magnitude of the lateral oscillation in $Y$ direction increases with decreasing $\dot{\gamma}_{\rm w}$. 
Thus for further decreasing $\dot{\gamma}_{\rm w}$, the magnitude of the lateral oscillation exceeds the width of the channel, resulting in the particle bouncing between the two walls as shown in fig.~\ref{fig:PoiseuilleFlow_Wall_parallel}(b) for $\alpha=0.1$ and $\dot{\gamma}_{\rm w}=0.01$. 
Again, the active velocity oscillates around the upstream direction and the particle exhibits a net upstream migration as shown in fig.~\ref{fig:PoiseuilleFlow_Wall_parallel}(b). 
This bouncing motion between two channel walls is observed at the grey pluses in fig.~\ref{fig:PoiseuilleFlow_Wall_parallel}(a). 

At the magenta down triangles in fig.~\ref{fig:PoiseuilleFlow_Wall_parallel}(a), a tumbling motion was obtained. 
The tumbling particles exhibit full rotation of the particle shape due to the passive rotation by the flow vorticity $\mathsf{W}$ as depicted by the sky blue (light grey) lines in fig.~\ref{fig:PoiseuilleFlow_Wall_parallel}(e) for $\alpha=0.9$ and $\dot{\gamma}_{\rm w}=0.06$. 
However, due to the deformability, the particle shape may fail to make a full rotation. 
In fact, for the parameters denoted by the pink up triangles in fig.~\ref{fig:PoiseuilleFlow_Wall_parallel}(a), the particles bounce against the channel wall with the shape oscillating as depicted by the sky blue lines in fig.~\ref{fig:PoiseuilleFlow_Wall_parallel}(c) for $\alpha=0.1$ and $\dot{\gamma}_{\rm w}=0.02$. 
In this case, the longitudinal axis of the elliptical particle shape oscillates around $\theta=0$, as shown in fig.~\ref{fig:PoiseuilleFlow_Wall_parallel}(c).
In contrast, at the purple diamonds in fig.~\ref{fig:PoiseuilleFlow_Wall_parallel}(a), the particles bounce against the wall with the shape oscillating around the diagonal direction, as depicted by the sky blue lines in fig.~\ref{fig:PoiseuilleFlow_Wall_parallel}(d) for $\alpha=0.1$ and $\dot{\gamma}_{\rm w}=0.04$. 
In this case, the particles are stretched in the direction $\theta = -\pi/4$ ($\pi/4$) for $y>0$ ($y<0$) by the external flow through $\mathsf{A}$ defined in eq.~\eqref{eq:A}. 
Therefore, the angle $\theta$ oscillates around $\theta = -\pi/4$ ($\pi/4$) for $y>0$ ($y<0$). 
Note that, although the trajectories depicted by the sky blue (light grey) lines in fig~\ref{fig:PoiseuilleFlow_Wall_parallel}(c,d,e) are all bouncing against the upper channel wall placed at $Y=+1$, there also exist their counterparts bouncing against the lower channel wall at $Y=-1$. 

When the self-propulsion is large, $\alpha > \alpha_c$, the particle undergoes a circular motion in the absence of the external flow. 
Then, if the external flow is turned on, it exhibits a cycloidal motion as shown in fig.~\ref{fig:PoiseuilleFlow_Wall_parallel}(e) for $\alpha=0.9$ and $\dot{\gamma}_{\rm w}=0.06$. 
The turquoise dots in fig.~\ref{fig:PoiseuilleFlow_Wall_parallel}(a) represent the parameters for which the cycloidal motion was obtained. 
This cycloidal motion appears as a result of the superposition of the circular motion and the passive advection due to the Poiseuille flow, in the same manner as the cycloidal solution obtained in the case of a linear shear flow~\cite{Tarama2013DynamicsJCP}. 
Interestingly, in the Poiseille flow, the particles undergoing the cycloidal motion drift in the vertical direction until they reach the preferred $Y$ position (i.e., a preferred local shear rate), which gets closer to the centerline as $\dot{\gamma}_{\rm w}$ increases. 
Finally, for a large $\dot{\gamma}_{\rm w}$, the particle exhibits a bouncing motion against the wall with excursion to the middle of the channel as depicted in fig.~\ref{fig:PoiseuilleFlow_Wall_parallel}(f) for $\alpha=0.1$ and $\dot{\gamma}_{\rm w}=0.4$. 
This motion is observed for the orange crosses in fig.~\ref{fig:PoiseuilleFlow_Wall_parallel}(a). 
Although only the solution bouncing against the upper channel wall is displayed in fig~\ref{fig:PoiseuilleFlow_Wall_parallel}(f), there also exists the one bouncing against the lower channel wall at $Y=-1$.

\subsection{Perpendicular particles} \label{section:perpendicular_poiseuille_flow}

Now we consider the perpendicular particles, which are self-propelling perpendicularly to the longitudinal axis of the elliptical shape deformation ($a_1=b_1=-1$). 
Unlike the parallel particles, the dynamics of the perpendicular particles are very different from the case of the rigid active particles discussed in sec.~\ref{section:rigid_poiseuille_flow}. 
In particular, the swinging motion is not observed for the perpendicular particles. 
Instead, a straight motion along a streamline far from the channel walls is observed, as we explain below. 

\begin{figure*}[tb]
\begin{center}
 \includegraphics[width=\textwidth]{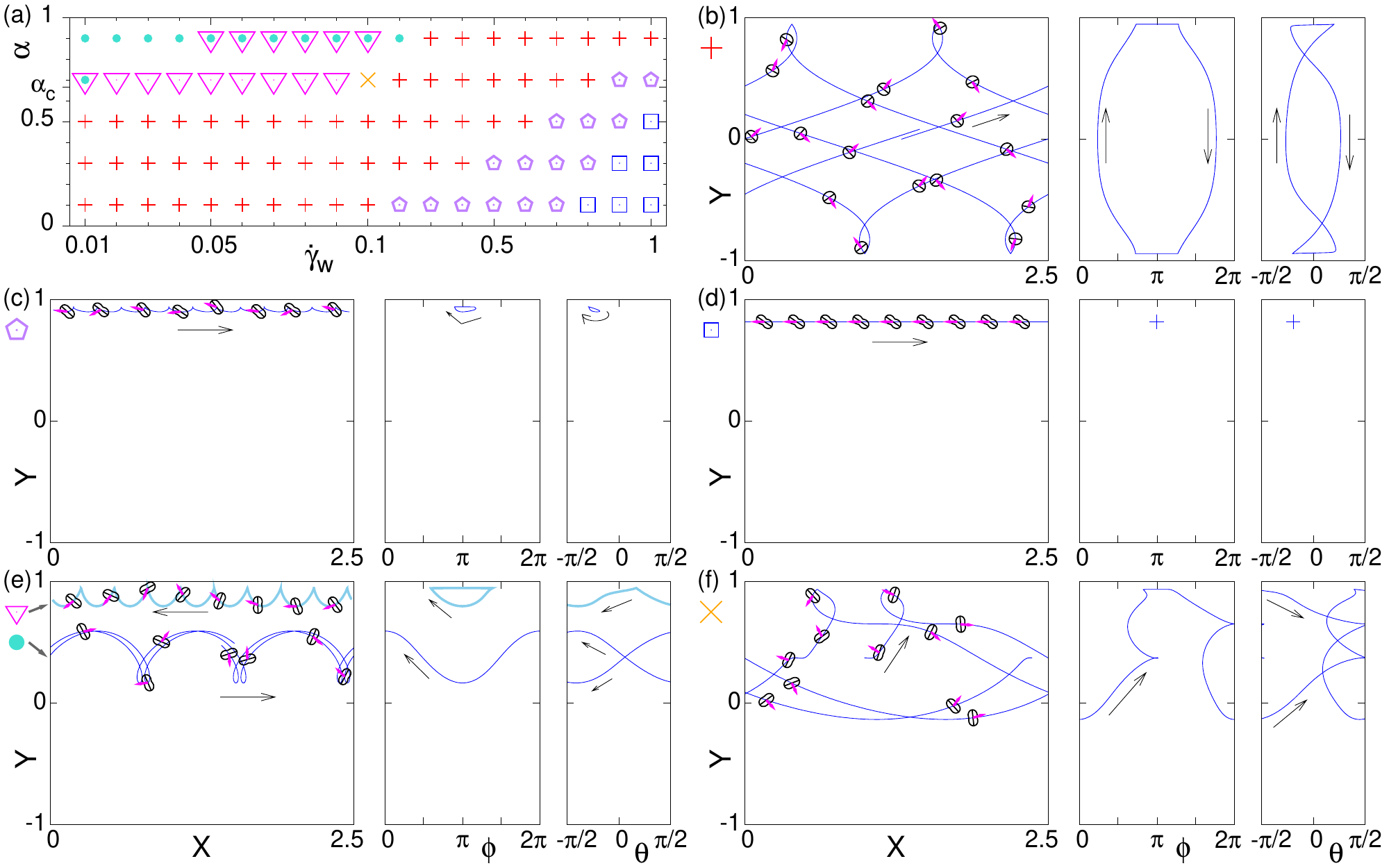}
\caption{(Colour online)
Dynamics of perpendicular particles ($a_1=b_1=-1$) in the Poiseuille flow. 
(a) Dynamical phase diagram for varying $\alpha$ and $\dot{\gamma}_{\rm w}$. 
(b)--(f) Characteristic trajectories of each dynamical motion in real (left), $Y$-$\phi$ (middle), and $Y$-$\theta$ (right) spaces. 
(b) Bouncing between the two channel walls for $\alpha=0.1$ and $\dot{\gamma}_{\rm w}=0.03$
(c) Bouncing motion against one of the two walls for $\alpha=0.3$ and $\dot{\gamma}_{\rm w}=0.5$;
(d) Straight motion far from the channel wall for $\alpha=0.3$ and $\dot{\gamma}_{\rm w}=1$;
(e) Cycloidal motion [blue (dark grey) lines] and tumbling motion at the wall [sky blue (light grey) lines] for $\alpha=0.9$ and $\dot{\gamma}_{\rm w}=0.06$;
(f) Bouncing against the wall with an excursion to the middle of the channel for $\alpha=0.7$ and $\dot{\gamma}_{\rm w}=0.1$.
See also the caption of fig.~\ref{fig:PoiseuilleFlow_Wall_parallel} for more explanation. 
}%
\label{fig:PoiseuilleFlow_Wall_perpendicular}
\end{center}
\end{figure*}

First, for $0<\alpha<\alpha_c$, the particles bounce between the two channel walls as depicted in fig.~\ref{fig:PoiseuilleFlow_Wall_perpendicular}(b) for $\alpha=0.1$ and $\dot{\gamma}_{\rm w}=0.03$. 
This motion is obtained at the red pluses in fig.~\ref{fig:PoiseuilleFlow_Wall_perpendicular}(a). 
For larger $\dot{\gamma}_{\rm w}$, the particles undergo a bouncing motion against only one of the two channel walls. 
This bouncing motion is observed at the purple pentagons in fig.~\ref{fig:PoiseuilleFlow_Wall_perpendicular}(a) and the typical trajectories are displayed in fig.~\ref{fig:PoiseuilleFlow_Wall_perpendicular}(c) for $\alpha=0.3$ and $\dot{\gamma}_{\rm w}=0.5$. 
The active velocity undergoes oscillations around the upstream direction and the particle shape oscillates around the stretching direction due to the external flow, i.e., around $\theta = -\pi/4$ ($\pi/4$) for $y>0$ ($y<0$). 
The magnitude of the oscillation decreases with increasing $\dot{\gamma}_{\rm w}$. 

Upon further increasing $\dot{\gamma}_{\rm w}$, the active velocity and the shape deformation cease oscillating. 
Then, the particles exhibit a straight motion along a streamline as depicted in fig.~\ref{fig:PoiseuilleFlow_Wall_perpendicular}(d) for $\alpha=0.3$ and $\dot{\gamma}_{\rm w}=1$. 
It is emphasized that this straight motion occurs far from the channel walls and, for the same parameters, the particles starting from different initial conditions end up at the same distance from the channel walls, which increases with $\dot{\gamma}_{\rm w}$. 
This indicates that the particles have a preferred local shear rate. 
Note that, for the straight motion, the active velocity points upstream, i.e., $\phi=\pi$, and the longitudinal axis of the elliptical deformation takes a value slightly larger (smaller) than $\theta = -\pi/4$ ($\theta=\pi/4$) for $y>0$ ($y<0$), where the torque from the following two contributions balance. 
The first one is the vorticity of the external flow $\mathsf{W}$, which rotates the particle in the counterclockwise (clockwise) direction for $y>0$ ($y<0$). 
The second torque comes from the stretching due to the external flow $\mathsf{A}$ that tends to align the shape to $\theta = -\pi/4$ ($\theta=\pi/4$) for $y>0$ ($y<0$) so that it rotates the particle shape in the clockwise (counterclockwise) direction if $\theta$ is slightly larger (smaller) than $-\pi/4$ ($+\pi/4$) for $y>0$ ($y<0$). 
The straight motion along a streamline far from the channel wall is obtained for the blue squares in fig.~\ref{fig:PoiseuilleFlow_Wall_perpendicular}(a). 

For $\alpha>\alpha_c$, the particles exhibit a cycloidal motion as displayed by the blue lines in fig.~\ref{fig:PoiseuilleFlow_Wall_perpendicular}(e) for $\alpha=0.9$ and $\dot{\gamma}_{\rm w}=0.06$. 
As in the case of the parallel particles, the particles undergoing the cycloidal motion drift vertically until they reach the $Y$ position with the preferred local shear rate. 
The cycloidal motion is obtained at the turquoise dots in fig.~\ref{fig:PoiseuilleFlow_Wall_perpendicular}(a), while the magenta down triangles denote a tumbling motion. 
The typical trajectories of the tumbling motion are plotted by the sky blue lines in fig.~\ref{fig:PoiseuilleFlow_Wall_perpendicular}(e) for $\alpha=0.9$ and $\dot{\gamma}_{\rm w}=0.06$. 
At the orange cross in fig.~\ref{fig:PoiseuilleFlow_Wall_perpendicular}(a), the particle exhibits a bouncing against the channel wall with an excursion to the middle of the channel as depicted in fig.~\ref{fig:PoiseuilleFlow_Wall_perpendicular}(f) for $\alpha=0.7$ and $\dot{\gamma}_{\rm w}=0.1$.

\section{Overdamped limit} \label{section:overdamped_poiseuille_flow}

So far, we show the dynamics of active rigid particles and active deformable particles in an external Poiseuille flow. 
In the case of rigid elliptical active particles in a Poiseuille flow, the dynamics are qualitatively the same for the parallel and perpendicular configurations and a swinging motion around the centerline is a major solution, as discussed in Sec.~\ref{section:rigid_poiseuille_flow}~\cite{Zottl2012Nonlinear,Zottl2013Periodic}. 
Interestingly, however, in the case of active deformable particles, the dynamics are different for the parallel and perpendicular particles. 
Although active deformable particles that tend to self-propel in the parallel direction with respect to the longitudinal axis of the elliptical shape deformation exhibit a swinging motion around the centerline, the swinging motion is not observed for the active deformable particles self-propelling perpendicularly. 
Now we ask ourselves what the origin of the difference in the dynamics between the active deformable particles self-propelling in the parallel and perpendicular directions is. 

Comparing the equations of motion for the deformable active particles \eqref{eq:dx}--\eqref{eq:dtheta} to those for the rigid active particles, eqs.~\eqref{eq:dx}, \eqref{eq:dy}, and \eqref{eq:dphi=dtheta}, there are two major differences. 
One is obviously the particle deformability. 
The other is the time-derivative term of the active velocity in eq.~\eqref{eq:dv}. 
Note that this inertia-like term is also derived from the Stokes equation for an active droplet on the interface of which a chemical reaction takes place to change the local surface tension~\cite{Yabunaka2012Self-propelled,Yoshinaga2014Spontaneous}. 
In this section, we investigate which of these two difference is relevant for the difference in dynamics, especially the existence of the swinging motion. 
For this purpose, we carried out numerical simulations of eqs.~\eqref{eq:dx}--\eqref{eq:dtheta} without the time-derivative term in eq.~\eqref{eq:dv}, which corresponds to the overdamped limit. 

In this limit, eqs.~\eqref{eq:dr_i} and \eqref{eq:dv_i}, and thus, eqs.~\eqref{eq:dx}--\eqref{eq:dphi}, are replaced by
\begin{align}
&\frac{d r_i}{dt} = v_i +u_i + f_i, 
 \label{eq:overdamped:dr_i}\\
&v = \big( \alpha -a_1 s \cos2\psi \big)^{1/2}
 \label{eq:overdamped:v}\\
&\frac{d \phi}{d t}
 = -a_1 s \sin2\psi + \frac{\dot{\gamma}}{2} Y, 
 \label{eq:overdamped:phi}
\end{align}
with eq.~\eqref{eq:v}, whereas the equation for the deformation, eq.~\eqref{eq:dS_ij}, or its equivalent eqs.~\eqref{eq:ds} and \eqref{eq:dtheta}, is unaltered.  
Then the equations to solve here are eqs.~\eqref{eq:overdamped:dr_i}--\eqref{eq:overdamped:phi} together with eqs.~\eqref{eq:ds}, and \eqref{eq:dtheta}. 

\begin{figure}[tb]
\begin{center}
 \includegraphics[width=\columnwidth]{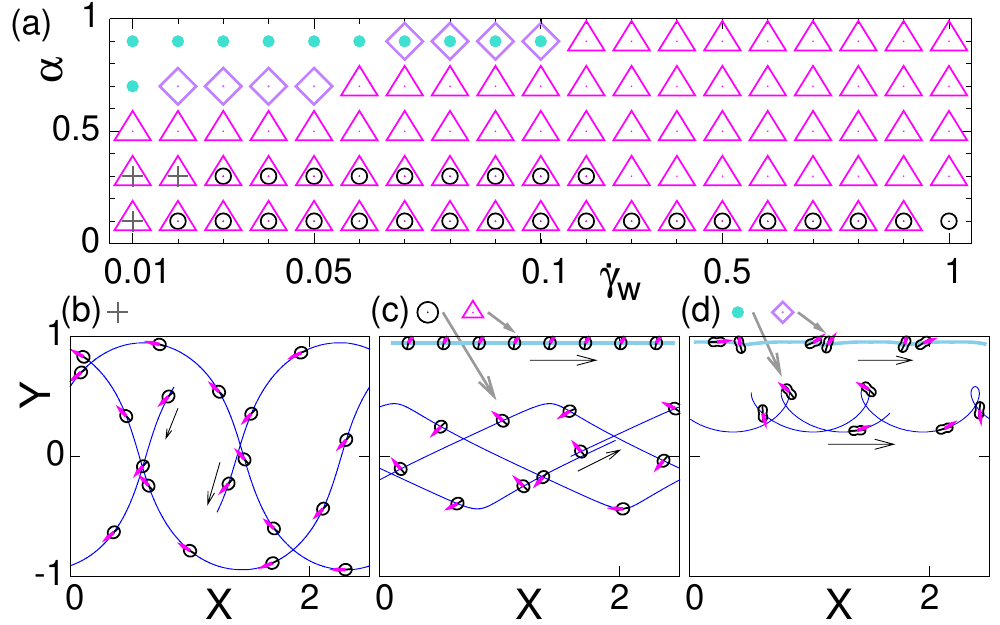}
\caption{(Colour online)
Overdamped dynamics of the parallel particles ($a_1=b_1=1$) with eq.~\eqref{eq:overdamped:v} in the Poiseuille flow. 
Different dynamical states are denoted by different symbols in panel (a), which are plotted in panels (b)--(f) consistently. 
(a) Dynamical phase diagram for varying $\alpha$ and $\dot{\gamma}_{\rm w}$. 
The superposition of the symbols indicate coexistence. 
Note that the scale of the $x$ axis is different for $\dot{\gamma}_{\rm w}<0.1$ and for $\dot{\gamma}_{\rm w}>0.1$. 
(b)--(d) Characteristic trajectory of each dynamical motion in real space. 
(b) Bouncing motion between the two channel walls for $\alpha=0.1$ and $\dot{\gamma}_{\rm w}=0.01$;
(c) Swinging motion [blue (dark grey) line] and straight motion along the channel wall [sky blue (light grey) line] for $\alpha=0.1$ and $\dot{\gamma}_{\rm w}=0.05$;
(d) Cycloidal motion [blue (dark grey) line] and oscillating motion along the wall [sky blue (light grey) line] for $\alpha=0.9$ and $\dot{\gamma}_{\rm w}=0.07$. 
In panels (b)--(d), some particle silhouettes are superposed onto the real space trajectories with the magenta arrows and the black bars indicating the direction of the self-propulsion $\bm{v}$ and the longitudinal axis of the elliptical deformation, respectively. 
Black arrows indicate the time-evolution direction. 
The grey arrows by the symbols in panels (c) and (d) point to the corresponding trajectories.  
}%
\label{fig:PoiseuilleFlow_VofS_Wall_parallel}
\end{center}
\end{figure}

\begin{figure}[tb]
\begin{center}
 \includegraphics[width=\columnwidth]{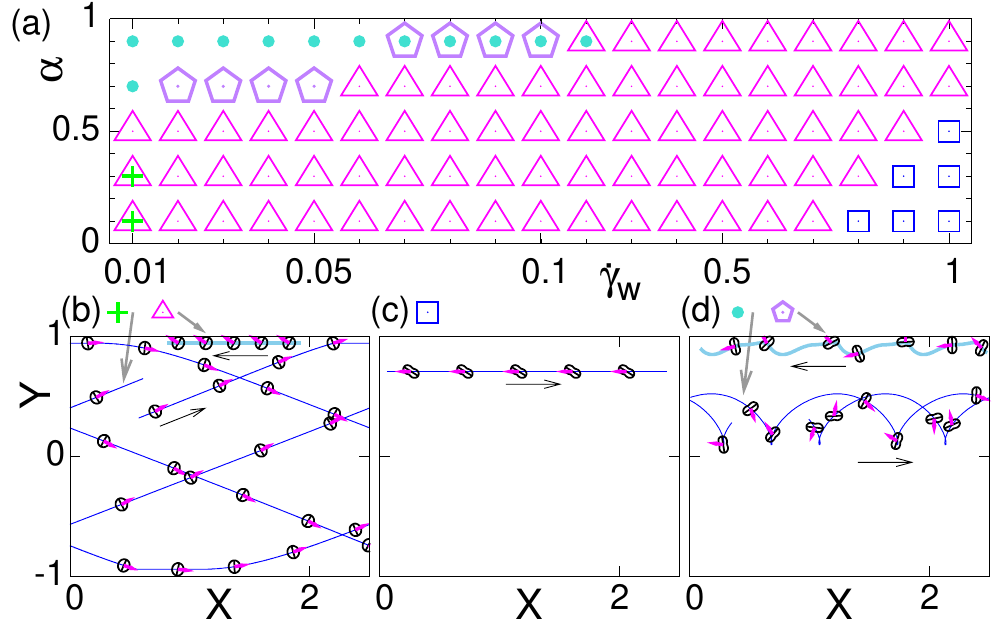}
\caption{(Colour online)
Overdamped dynamics of the perpendicular particles ($a_1=b_1=-1$) with eq.~\eqref{eq:overdamped:v} in the Poiseuille flow. 
(a) Dynamical phase diagram for varying $\alpha$ and $\dot{\gamma}_{\rm w}$. 
(b)--(d) Characteristic trajectory of each dynamical motion in real space. 
(b) Bouncing motion between the two channel walls [blue (dark grey) line] and straight motion along the wall [sky blue (light grey) line] for $\alpha=0.1$ and $\dot{\gamma}_{\rm w}=0.01$;
(c) Straight motion far from the channel wall for $\alpha=0.1$ and $\dot{\gamma}_{\rm w}=1$;
(d) Cycloidal motion [blue (dark grey) line] and oscillating motion along the wall [sky blue (light grey) line] for $\alpha=0.9$ and $\dot{\gamma}_{\rm w}=0.07$. 
See also the caption of fig.~\ref{fig:PoiseuilleFlow_VofS_Wall_parallel}. 
}%
\label{fig:PoiseuilleFlow_VofS_Wall_perpendicular}
\end{center}
\end{figure}

Now we show the results of the overdamped dynamics of deformable active particles. 
Again, we distinguish the parallel particles, which tend to self-propel parallel to the longitudinal axis of the elliptical shape deformation, and the perpendicular particles, self-propelling perpendicularly.  
The dynamics of the overdamped dynamics of the parallel ($a_1=b_1=1$) and perpendicular particles ($a_1=b_1=-1$) are summarized in figs.~\ref{fig:PoiseuilleFlow_VofS_Wall_parallel} and \ref{fig:PoiseuilleFlow_VofS_Wall_perpendicular}, respectively. 
In both figures, the phase diagram and characteristic real-space trajectories of each dynamical state are depicted in panel (a) and in panels (b)--(d), respectively. 
In panels (b)--(d), a periodic boundary condition with the window width $X=2.5$ is applied in $x$ direction. 
The resulting dynamical motions and the phase diagrams are similar to those of the underdamped case discussed in sec.~\ref{section:deformable_poiseuille_flow}, except for the bouncing motion against the channel walls. 
In the following, we discuss the parallel and perpendicular particles one by one. 

First, the parallel particles exhibit a swinging motion around the center line for the parameters plotted by the black circles in fig.~\ref{fig:PoiseuilleFlow_VofS_Wall_parallel}(a). 
The particles move far from the confinement as depicted by the blue (dark grey) line in fig.~\ref{fig:PoiseuilleFlow_VofS_Wall_parallel}(c) for $\alpha=0.1$ and $\dot{\gamma}_{\rm w}=0.05$. 
Note that an upstream swinging motion is found for a sufficiently small $\dot{\gamma}_{\rm w}$. 
For even smaller $\dot{\gamma}_{\rm w}$, a bouncing motion between the two channel walls is observed, as displayed in fig.~\ref{fig:PoiseuilleFlow_VofS_Wall_parallel}(b) for $\alpha=0.1$ and $\dot{\gamma}_{\rm w}=0.01$. 
This motion is obtained for the parameters denoted by the grey pluses in fig.~\ref{fig:PoiseuilleFlow_VofS_Wall_parallel}(a). 
In both cases, the active velocity oscillates around the upstream direction. 
In fact, these two solutions correspond to those in the underdamped case discussed in Sec.~\ref{section:parallel_poiseuille_flow} (see fig.~\ref{fig:PoiseuilleFlow_Wall_parallel}). 

At the magenta triangles in fig.~\ref{fig:PoiseuilleFlow_VofS_Wall_parallel}(a), the particles undergo a straight motion along the channel wall. 
This is because the time-evolution equation is now overdamped so that once the particles go into the wall, it is difficult to escape from it. 
The straight motion along the wall is depicted by the sky blue (light grey) line in fig.~\ref{fig:PoiseuilleFlow_VofS_Wall_parallel}(c) for $\alpha=0.1$ and $\dot{\gamma}_{\rm w}=0.05$. 

For a large $\alpha$, the particles undergoing a circular motion in the absence of the external flow, exhibit a cycloidal motion at the turquoise dots in fig~\ref{fig:PoiseuilleFlow_VofS_Wall_parallel}(a). 
A typical trajectory of the cycloidal motion is displayed by the blue (dark grey) line in fig.~\ref{fig:PoiseuilleFlow_VofS_Wall_parallel}(d) for $\alpha=0.9$ and $\dot{\gamma}_{\rm w}=0.07$. 
The cycloidal motion obtained here is the same solution as that of fig.~\ref{fig:PoiseuilleFlow_Wall_parallel}(e) where the dynamics is underdamped. 
At the purple diamonds in fig.~\ref{fig:PoiseuilleFlow_VofS_Wall_parallel}(a) an oscillatory motion along the channel wall is observed, as depicted by the sky blue (light grey) line in fig.~\ref{fig:PoiseuilleFlow_VofS_Wall_parallel}(d) for $\alpha=0.9$ and $\dot{\gamma}_{\rm w}=0.07$.

Next we consider the overdamped dynamics of the active deformable particles self-propelling perpendicularly to the elliptical deformation. 
For these particles, a swinging motion is not observed, as in the underdamped case in sec.~\ref{section:perpendicular_poiseuille_flow}. 

For the magenta triangles in fig.~\ref{fig:PoiseuilleFlow_VofS_Wall_perpendicular}, they exhibit a straight motion along the channel wall as depicted by the sky blue (light grey) line in fig.~\ref{fig:PoiseuilleFlow_VofS_Wall_perpendicular}(b) for $\alpha=0.1$ and $\dot{\gamma}_{\rm w}=0.01$. 
For this straight motion, the active velocity $\bm{v}$ points upstream. 
In addition, at the green pluses in fig.~\ref{fig:PoiseuilleFlow_VofS_Wall_perpendicular}(a), the particles bounce between the two channel walls as depicted by the blue (dark grey) line in fig.~\ref{fig:PoiseuilleFlow_VofS_Wall_perpendicular}(b) for $\alpha=0.1$ and $\dot{\gamma}_{\rm w}=0.01$. 
Unlike the straight motion along the wall, the oscillating active velocity points towards the downstream direction. 
Therefore, this bouncing motion corresponds to a different solution from the bouncing motion between the two channel walls for the parallel case shown in fig.~\ref{fig:PoiseuilleFlow_VofS_Wall_parallel}(b). 
Note that the bouncing motion in fig.~\ref{fig:PoiseuilleFlow_VofS_Wall_parallel}(b) occurs as a consequence of the swinging parallel particle hitting the walls due to the large oscillation in the $Y$ direction, for which $\bm{v}$ tends to be upstream. 

For large $\dot{\gamma}_{\rm w}$, the particles exhibit a straight motion along a streamline far from the channel walls, as depicted in fig.~\ref{fig:PoiseuilleFlow_VofS_Wall_perpendicular}(c) for $\alpha=0.1$ and $\dot{\gamma}_{\rm w}=1$. 
This motion, obtained for the blue squares in fig.~\ref{fig:PoiseuilleFlow_VofS_Wall_perpendicular}(a), is the same solution as that observed in the underdamped case in fig.~\ref{fig:PoiseuilleFlow_Wall_perpendicular}(d). 

At the turquoise dots in fig.~\ref{fig:PoiseuilleFlow_VofS_Wall_perpendicular}(a), a cycloidal motion is observed, as in the underdamped case in fig.~\ref{fig:PoiseuilleFlow_Wall_perpendicular}. 
The purple pentagons in fig.~\ref{fig:PoiseuilleFlow_VofS_Wall_perpendicular}(a) denote an oscillatory motion along the channel wall. 
Typical trajectories of the cycloidal motion and the oscillatory motion along the wall are displayed by the blue (dark grey) line and by the sky blue (light grey) line, respectively, in fig.~\ref{fig:PoiseuilleFlow_VofS_Wall_perpendicular}(d) for $\alpha=0.9$ and $\dot{\gamma}_{\rm w}=0.07$.  

In conclusion, concerning the difference between the parallel and perpendicular particles, the overdamped equations described in this section reproduced the same dynamical motions far from the channel walls as in the underdamped case discussed in Sec.~\ref{section:deformable_poiseuille_flow}. 
Indeed, while the swinging motion around the centerline is observed for the parallel particles, the perpendicular particles exhibit the straight motion along a streamline far from the walls. 
Similar results are also obtained for the case of a constant active velocity, eq.~\eqref{eq:v0}, which is independent of the magnitude of the deformation (data not shown). 
Consequently, we conclude that the time-derivative term of the active velocity is not crucial for the different dynamics for the deformable active particles self-propelling parallel and perpendicularly to the elliptical shape. 
However, the deformability of the particle shape is the main cause of the difference in the dynamics. 
In fact, we numerically find that, if the deformability is omitted from eqs.~\eqref{eq:dr_i}, \eqref{eq:dv_i}, and \eqref{eq:dS_ij} and the magnitude of the deformation is fixed by eq.~\eqref{eq:s0}, similar dynamics to those discussed in Sec.~\ref{section:rigid_poiseuille_flow} are obtained (data not shown). 
Namely, the swinging motion around the centerline is observed both for the parallel and perpendicular cases.

\section{Discussion} \label{section:discussion}

In this paper, we investigated the dynamics of active deformable particles in an external Poiseuille flow. 
To make the analysis general, we employed the time-evolution equations that we derived previously based on symmetry considerations for the center-of-mass position and velocity and the elliptical shape deformation~\cite{Tarama2013DynamicsJCP}. 
The self-propulsion is included by introducing a supercritical pitchfork bifurcation structure in the equation for the active velocity. 
Due to the coupling between the active velocity and the shape deformation, it is known~\cite{Ohta2009Deformable,Tarama2013DynamicsJCP} that there is a bifurcation at $\alpha=\alpha_c$ 
 in the absence of an external flow field; 
The particles undergo a straight motion for a small self-propulsion $0<\alpha<\alpha_c$, and a circular motion for a large self-propulsion $\alpha>\alpha_c$. 

We first considered the limit of rigid active particles with a circular shape and an elliptical shape. 
In this limit, we showed that our equations were reduced to the model equations studied by Z\"ottl and Stark~\cite{Zottl2012Nonlinear,Zottl2013Periodic}. 
For both circular and elliptical rigid active particles, the time-evolution equations were solved analytically and the solutions representing a swinging motion around the centerline and a tumbling motion far from the centerline were obtained. 
For elliptical rigid active particles, we clarified that the ones self-propelling in the parallel direction with respect to the longitudinal axis of the elliptical shape and those self-propelling perpendicularly showed qualitatively the same dynamics.

In contrast, for deformable active particles in a Poiseuille flow, we numerically solved the equations of motion and found qualitatively different dynamics for the parallel particles, which tend to self-propel parallel to the longitudinal axis of the elliptical deformation, and for the perpendicular particles, that are self-propelling perpendicularly. 
Note that in the case of active deformable particles in a simple linear shear flow~\cite{Tarama2013DynamicsJCP}, there is a symmetry in the equations of motion, which makes the dynamics equivalent for the parallel and perpendicular particles~\cite{Tarama2017Dynamics}. 
However, such symmetry was not found for the present case of the Poiseuille flow. 

For a small self-propulsion $0<\alpha<\alpha_c$, the parallel particles exhibited the swinging motion around the centerline as well as a tumbling motion around the channel wall, and a bouncing motion against the wall under the external Poiseuille flow field. 
However, the perpendicular particles underwent a bouncing motion between the two channel walls for a small flow velocity and a straight motion along a streamline far from the channel walls for a large flow velocity. 
For an intermediate flow velocity, there was a parameter region where the particles bounced against one of the two channel walls. 

For a large self-propulsion $\alpha>\alpha_c$, for both parallel and perpendicular particles, a cycloidal motion parallel to the flow direction and a tumbling motion around the wall were obtained for a small flow velocity. 
For a large flow velocity, the parallel particles bounced against a channel wall with an excursion to the middle of the channel, while the bouncing motion between the two channel walls were obtained for the perpendicular particles.
For an even larger flow velocity, the perpendicular particles exhibited a bouncing motion against one of the two walls as well as a straight motion along a streamline far from the channel walls.

On the one hand, the swinging motion around the centerline was a solution for the active deformable particles self-propelling parallel to the elliptical deformation as in the case of the rigid active particles. 
On the other hand, unlike the rigid case, the swinging motion was not found for the active deformable particles self-propelling perpendicularly, which instead exhibited a straight motion along a streamline far from the channel walls. 
In order to clarify the origin of the difference between the parallel and perpendicular cases, we considered the overdamped limit of the model equations. 
We numerically confirmed that the dynamical motions far from the channel walls were unaltered, while those moving around the walls
 were mostly transformed into a straight motion along the wall. 
Therefore, we concluded that the difference was caused by the degree of freedom of the particle shape deformation.

In this study, the external flow profile is assumed to be prescribed and the effect of the particle self-propulsion on the surrounding flow is omitted. 
This effect is important for, e.g., the hydrodynamic interaction with other particles and with confinements. 
Still, our current simple approach provides a good approximation for a single particle moving far from the channel walls. 
Therefore,
although the trajectories that come to the close vicinity of the walls may be significantly modified from the ones presented here if the hydrodynamic interactions are taken into consideration,
 the dynamical states that stay far from the channel walls are likely to be unaltered. 
Thus, the solutions obtained in this study, such as the swinging motion around the centerline for the parallel particles and the straight motion along the streamline far from the channel walls for the perpendicular particles, are expected to be observed experimentally. 

The hydrodynamic interaction can be included either by solving the full hydrodynamic equations~\cite{Zottl2014Hydrodynamics,Oyama2016Purely} or by the Green's function method~\cite{Lopez2014Dynamics}, which is usually more complicated. 
In fact, due to the hydrodynamic interactions, an anomalous change in viscosity is know for suspensions of bacteria~\cite{Hatwalne2004Rheology,Sokolov2009Reduction,Marcos2012Bacterial,Gachelin2013Non-Newtonian,Lopez2015Turning,Figueroa-Morales2015Living,Clement2016Bacterial}, which hardly change their shape. 
Therefore, as a future problem, we mention that it is interesting to see the impact of the shape deformation of the active particles on such rheological property.

\acknowledgements

We thank Hartmut~L\"owen for helpful discussions. 
John.~J.~Molina and Simon~K.~Schnyder are acknowledged for careful reading of the manuscript.

\end{document}